\documentclass[aps, twocolumn, superscriptaddress, showpacs, nofootinbib]{revtex4-1}

\usepackage{lipsum}
\usepackage[utf8]{inputenc}
\usepackage[T1]{fontenc}
\usepackage{ae,aecompl} 
\usepackage{graphicx}
\usepackage{amsmath}
\usepackage{color}
\usepackage{amssymb}
\usepackage{latexsym}
\usepackage{wasysym}
\usepackage{psfrag}
\usepackage{comment}
\usepackage{ifthen}
\usepackage[citecolor=blue,colorlinks=true]{hyperref}
\usepackage{longtable}
\usepackage{float}
\usepackage[utf8]{inputenc}
\usepackage{lineno}

\restylefloat{table}

\newcommand{\macro}[1]{\textcolor{red}{#1}} 

\newcommand{\Msun}{\ensuremath{\mathrm{M}_\odot}}

\newcommand\OBSEVENTTIME{\macro{09:50:45}}

\newcommand\BURSTEVENTIFAR{\macro{\ensuremath{22\,500}}}
\newcommand\BURSTBCKLIVETIME{\macro{\ensuremath{67\,400}}}

\newcommand{\RECONSTRUCTIONOVERLAP}{\macro{\ensuremath{94\%}}}

\newcommand{\PESKYFIFTYNOCALIB}{\macro{\ensuremath{48~\mathrm{deg^2}}}} 
\newcommand{\PESKYFIFTY}{\macro{\ensuremath{150~\mathrm{deg^2}}}} 
\newcommand{\PESKYNINTYNOCALIB}{\macro{\ensuremath{150~\mathrm{deg^2}}}} 
\newcommand{\PESKYNINTY}{\macro{\ensuremath{610~\mathrm{deg^2}}}} 
\newcommand{\MCobsCOMPACT}{\macro{\ensuremath{30_{-2}^{+2}}}} 
\newcommand{\MTOTobsCOMPACT}{\macro{\ensuremath{71_{-4}^{+5}}}} 
\newcommand{\MASSRATIOCOMPACT}{\macro{\ensuremath{0.82_{-0.20}^{+0.17}}}} 
\newcommand{\MTOTobsRANGE}{\macro{\ensuremath{66\text{--}75}}} 
\newcommand{\MCobsRANGE}{\macro{\ensuremath{29\text{--}33}}} 

\renewcommand{\macro}[1]{#1} 

\newcommand{\makevisible}[1]{\textbf{#1}}
\newcommand{\switch}[1]{%
  \ifthenelse{\equal{#1}{0}}{\renewcommand{\makevisible}[1]{}}{}}
\switch{0} 

\def\nr#1{numerical relativity#1
  (NR#1)\gdef\nr{NR}}

\begin{document}

\title{Observing gravitational-wave transient GW150914 with minimal assumptions}




\author{%
B.~P.~Abbott,$^{1}$  
R.~Abbott,$^{1}$  
T.~D.~Abbott,$^{2}$  
M.~R.~Abernathy,$^{1}$  
F.~Acernese,$^{3,4}$
K.~Ackley,$^{5}$  
C.~Adams,$^{6}$  
T.~Adams,$^{7}$
P.~Addesso,$^{3}$  
R.~X.~Adhikari,$^{1}$
% ...
\\
\medskip
(LIGO Scientific Collaboration and Virgo Collaboration)
\\
\medskip
M.~Clark,$^{63}$ 
R.~Haas,$^{29}$  
J.~Healy,$^{112}$    
I.~Hinder,$^{29}$ 
M.~Kinsey,$^{63}$ 
P.~Laguna$^{63}$ 
\\
\medskip
{{}$^{\dag}$Deceased, May 2015. {}$^{\ddag}$Deceased, March 2015. }%
}\noaffiliation
\affiliation {LIGO, California Institute of Technology, Pasadena, CA 91125, USA }
\affiliation {Louisiana State University, Baton Rouge, LA 70803, USA }
\affiliation {Universit\`a di Salerno, Fisciano, I-84084 Salerno, Italy }
\affiliation {INFN, Sezione di Napoli, Complesso Universitario di Monte S.Angelo, I-80126 Napoli, Italy }
\affiliation {University of Florida, Gainesville, FL 32611, USA }
\affiliation {LIGO Livingston Observatory, Livingston, LA 70754, USA }
\affiliation {Laboratoire d'Annecy-le-Vieux de Physique des Particules (LAPP), Universit\'e Savoie Mont Blanc, CNRS/IN2P3, F-74941 Annecy-le-Vieux, France }
\affiliation {Albert-Einstein-Institut, Max-Planck-Institut f\"ur Gravi\-ta\-tions\-physik, D-30167 Hannover, Germany }
\affiliation {Nikhef, Science Park, 1098 XG Amsterdam, Netherlands }
\affiliation {LIGO, Massachusetts Institute of Technology, Cambridge, MA 02139, USA }
\affiliation {Instituto Nacional de Pesquisas Espaciais, 12227-010 S\~{a}o Jos\'{e} dos Campos, S\~{a}o Paulo, Brazil }
\affiliation {INFN, Gran Sasso Science Institute, I-67100 L'Aquila, Italy }
\affiliation {INFN, Sezione di Roma Tor Vergata, I-00133 Roma, Italy }
\affiliation {Inter-University Centre for Astronomy and Astrophysics, Pune 411007, India }
\affiliation {International Centre for Theoretical Sciences, Tata Institute of Fundamental Research, Bangalore 560012, India }
\affiliation {University of Wisconsin-Milwaukee, Milwaukee, WI 53201, USA }
\affiliation {Leibniz Universit\"at Hannover, D-30167 Hannover, Germany }
\affiliation {Universit\`a di Pisa, I-56127 Pisa, Italy }
\affiliation {INFN, Sezione di Pisa, I-56127 Pisa, Italy }
\affiliation {Australian National University, Canberra, Australian Capital Territory 0200, Australia }
\affiliation {The University of Mississippi, University, MS 38677, USA }
\affiliation {California State University Fullerton, Fullerton, CA 92831, USA }
\affiliation {LAL, Universit\'e Paris-Sud, CNRS/IN2P3, Universit\'e Paris-Saclay, 91400 Orsay, France }
\affiliation {Chennai Mathematical Institute, Chennai 603103, India }
\affiliation {Universit\`a di Roma Tor Vergata, I-00133 Roma, Italy }
\affiliation {University of Southampton, Southampton SO17 1BJ, United Kingdom }
\affiliation {Universit\"at Hamburg, D-22761 Hamburg, Germany }
\affiliation {INFN, Sezione di Roma, I-00185 Roma, Italy }
\affiliation {Albert-Einstein-Institut, Max-Planck-Institut f\"ur Gravitations\-physik, D-14476 Potsdam-Golm, Germany }
\affiliation {APC, AstroParticule et Cosmologie, Universit\'e Paris Diderot, CNRS/IN2P3, CEA/Irfu, Observatoire de Paris, Sorbonne Paris Cit\'e, F-75205 Paris Cedex 13, France }
\affiliation {Montana State University, Bozeman, MT 59717, USA }
\affiliation {Universit\`a di Perugia, I-06123 Perugia, Italy }
\affiliation {INFN, Sezione di Perugia, I-06123 Perugia, Italy }
\affiliation {European Gravitational Observatory (EGO), I-56021 Cascina, Pisa, Italy }
\affiliation {Syracuse University, Syracuse, NY 13244, USA }
\affiliation {SUPA, University of Glasgow, Glasgow G12 8QQ, United Kingdom }
\affiliation {LIGO Hanford Observatory, Richland, WA 99352, USA }
\affiliation {Wigner RCP, RMKI, H-1121 Budapest, Konkoly Thege Mikl\'os \'ut 29-33, Hungary }
\affiliation {Columbia University, New York, NY 10027, USA }
\affiliation {Stanford University, Stanford, CA 94305, USA }
\affiliation {Universit\`a di Padova, Dipartimento di Fisica e Astronomia, I-35131 Padova, Italy }
\affiliation {INFN, Sezione di Padova, I-35131 Padova, Italy }
\affiliation {CAMK-PAN, 00-716 Warsaw, Poland }
\affiliation {University of Birmingham, Birmingham B15 2TT, United Kingdom }
\affiliation {Universit\`a degli Studi di Genova, I-16146 Genova, Italy }
\affiliation {INFN, Sezione di Genova, I-16146 Genova, Italy }
\affiliation {RRCAT, Indore MP 452013, India }
\affiliation {Faculty of Physics, Lomonosov Moscow State University, Moscow 119991, Russia }
\affiliation {SUPA, University of the West of Scotland, Paisley PA1 2BE, United Kingdom }
\affiliation {University of Western Australia, Crawley, Western Australia 6009, Australia }
\affiliation {Department of Astrophysics/IMAPP, Radboud University Nijmegen, 6500 GL Nijmegen, Netherlands }
\affiliation {Artemis, Universit\'e C\^ote d'Azur, CNRS, Observatoire C\^ote d'Azur, CS 34229, Nice cedex 4, France }
\affiliation {MTA E\"otv\"os University, ``Lendulet'' Astrophysics Research Group, Budapest 1117, Hungary }
\affiliation {Institut de Physique de Rennes, CNRS, Universit\'e de Rennes 1, F-35042 Rennes, France }
\affiliation {Washington State University, Pullman, WA 99164, USA }
\affiliation {Universit\`a degli Studi di Urbino ``Carlo Bo,'' I-61029 Urbino, Italy }
\affiliation {INFN, Sezione di Firenze, I-50019 Sesto Fiorentino, Firenze, Italy }
\affiliation {University of Oregon, Eugene, OR 97403, USA }
\affiliation {Laboratoire Kastler Brossel, UPMC-Sorbonne Universit\'es, CNRS, ENS-PSL Research University, Coll\`ege de France, F-75005 Paris, France }
\affiliation {Astronomical Observatory Warsaw University, 00-478 Warsaw, Poland }
\affiliation {VU University Amsterdam, 1081 HV Amsterdam, Netherlands }
\affiliation {University of Maryland, College Park, MD 20742, USA }
\affiliation {Center for Relativistic Astrophysics and School of Physics, Georgia Institute of Technology, Atlanta, GA 30332, USA }
\affiliation {Institut Lumi\`{e}re Mati\`{e}re, Universit\'{e} de Lyon, Universit\'{e} Claude Bernard Lyon 1, UMR CNRS 5306, 69622 Villeurbanne, France }
\affiliation {Laboratoire des Mat\'eriaux Avanc\'es (LMA), IN2P3/CNRS, Universit\'e de Lyon, F-69622 Villeurbanne, Lyon, France }
\affiliation {Universitat de les Illes Balears, IAC3---IEEC, E-07122 Palma de Mallorca, Spain }
\affiliation {Universit\`a di Napoli ``Federico II,'' Complesso Universitario di Monte S.Angelo, I-80126 Napoli, Italy }
\affiliation {NASA/Goddard Space Flight Center, Greenbelt, MD 20771, USA }
\affiliation {Canadian Institute for Theoretical Astrophysics, University of Toronto, Toronto, Ontario M5S 3H8, Canada }
\affiliation {Tsinghua University, Beijing 100084, China }
\affiliation {Texas Tech University, Lubbock, TX 79409, USA }
\affiliation {The Pennsylvania State University, University Park, PA 16802, USA }
\affiliation {National Tsing Hua University, Hsinchu City, 30013 Taiwan, Republic of China }
\affiliation {Charles Sturt University, Wagga Wagga, New South Wales 2678, Australia }
\affiliation {University of Chicago, Chicago, IL 60637, USA }
\affiliation {Caltech CaRT, Pasadena, CA 91125, USA }
\affiliation {Korea Institute of Science and Technology Information, Daejeon 305-806, Korea }
\affiliation {Carleton College, Northfield, MN 55057, USA }
\affiliation {Universit\`a di Roma ``La Sapienza,'' I-00185 Roma, Italy }
\affiliation {University of Brussels, Brussels 1050, Belgium }
\affiliation {Sonoma State University, Rohnert Park, CA 94928, USA }
\affiliation {Northwestern University, Evanston, IL 60208, USA }
\affiliation {University of Minnesota, Minneapolis, MN 55455, USA }
\affiliation {The University of Melbourne, Parkville, Victoria 3010, Australia }
\affiliation {The University of Texas Rio Grande Valley, Brownsville, TX 78520, USA }
\affiliation {The University of Sheffield, Sheffield S10 2TN, United Kingdom }
\affiliation {University of Sannio at Benevento, I-82100 Benevento, Italy and INFN, Sezione di Napoli, I-80100 Napoli, Italy }
\affiliation {Montclair State University, Montclair, NJ 07043, USA }
\affiliation {Universit\`a di Trento, Dipartimento di Fisica, I-38123 Povo, Trento, Italy }
\affiliation {INFN, Trento Institute for Fundamental Physics and Applications, I-38123 Povo, Trento, Italy }
\affiliation {Cardiff University, Cardiff CF24 3AA, United Kingdom }
\affiliation {National Astronomical Observatory of Japan, 2-21-1 Osawa, Mitaka, Tokyo 181-8588, Japan }
\affiliation {School of Mathematics, University of Edinburgh, Edinburgh EH9 3FD, United Kingdom }
\affiliation {Indian Institute of Technology, Gandhinagar Ahmedabad Gujarat 382424, India }
\affiliation {Institute for Plasma Research, Bhat, Gandhinagar 382428, India }
\affiliation {University of Szeged, D\'om t\'er 9, Szeged 6720, Hungary }
\affiliation {Embry-Riddle Aeronautical University, Prescott, AZ 86301, USA }
\affiliation {University of Michigan, Ann Arbor, MI 48109, USA }
\affiliation {Tata Institute of Fundamental Research, Mumbai 400005, India }
\affiliation {American University, Washington, D.C. 20016, USA }
\affiliation {University of Massachusetts-Amherst, Amherst, MA 01003, USA }
\affiliation {University of Adelaide, Adelaide, South Australia 5005, Australia }
\affiliation {West Virginia University, Morgantown, WV 26506, USA }
\affiliation {University of Bia{\l }ystok, 15-424 Bia{\l }ystok, Poland }
\affiliation {SUPA, University of Strathclyde, Glasgow G1 1XQ, United Kingdom }
\affiliation {IISER-TVM, CET Campus, Trivandrum Kerala 695016, India }
\affiliation {Institute of Applied Physics, Nizhny Novgorod, 603950, Russia }
\affiliation {Pusan National University, Busan 609-735, Korea }
\affiliation {Hanyang University, Seoul 133-791, Korea }
\affiliation {NCBJ, 05-400 \'Swierk-Otwock, Poland }
\affiliation {IM-PAN, 00-956 Warsaw, Poland }
\affiliation {Rochester Institute of Technology, Rochester, NY 14623, USA }
\affiliation {Monash University, Victoria 3800, Australia }
\affiliation {Seoul National University, Seoul 151-742, Korea }
\affiliation {University of Alabama in Huntsville, Huntsville, AL 35899, USA }
\affiliation {ESPCI, CNRS, F-75005 Paris, France }
\affiliation {Universit\`a di Camerino, Dipartimento di Fisica, I-62032 Camerino, Italy }
\affiliation {Southern University and A\&M College, Baton Rouge, LA 70813, USA }
\affiliation {College of William and Mary, Williamsburg, VA 23187, USA }
\affiliation {Instituto de F\'\i sica Te\'orica, University Estadual Paulista/ICTP South American Institute for Fundamental Research, S\~ao Paulo SP 01140-070, Brazil }
\affiliation {University of Cambridge, Cambridge CB2 1TN, United Kingdom }
\affiliation {IISER-Kolkata, Mohanpur, West Bengal 741252, India }
\affiliation {Rutherford Appleton Laboratory, HSIC, Chilton, Didcot, Oxon OX11 0QX, United Kingdom }
\affiliation {Whitman College, 345 Boyer Avenue, Walla Walla, WA 99362 USA }
\affiliation {National Institute for Mathematical Sciences, Daejeon 305-390, Korea }
\affiliation {Hobart and William Smith Colleges, Geneva, NY 14456, USA }
\affiliation {Janusz Gil Institute of Astronomy, University of Zielona G\'ora, 65-265 Zielona G\'ora, Poland }
\affiliation {Andrews University, Berrien Springs, MI 49104, USA }
\affiliation {Universit\`a di Siena, I-53100 Siena, Italy }
\affiliation {Trinity University, San Antonio, TX 78212, USA }
\affiliation {University of Washington, Seattle, WA 98195, USA }
\affiliation {Kenyon College, Gambier, OH 43022, USA }
\affiliation {Abilene Christian University, Abilene, TX 79699, USA }


\begin{abstract}

  The gravitational-wave signal GW150914 was first identified on Sept 14 2015  by searches for short-duration gravitational-wave transients.
  These searches identify time-correlated transients in multiple detectors with minimal assumptions about
  the signal morphology, allowing them to be sensitive to gravitational waves emitted by a wide range of
  sources including binary black-hole mergers.
  Over the observational period from September 12th to October 20th 2015, these
  transient searches were sensitive to binary black-hole mergers similar to GW150914 to an average distance of $\sim 600$\,Mpc.
In this paper, we describe the
  analyses that first detected GW150914 as well as the parameter estimation and waveform reconstruction techniques
  that initially identified GW150914 as the merger of two black holes. We find that the reconstructed waveform is
  consistent with the signal from a binary black-hole merger with a chirp mass of
  $\sim 30 ~\Msun$ and a total mass before merger of $\sim 70 ~\Msun$ in the detector frame.
  
\end{abstract}

\maketitle

\section{Introduction}

The newly upgraded Advanced LIGO observatories
\cite{TheLIGOScientific:2014jea, GW150914-DETECTORS},
with sites near Hanford, WA (H1) and
Livingston, LA (L1),
host the most sensitive gravitational-wave detectors ever built.
The observatories use kilometer-scale Michelson interferometers that are
designed to detect
small, traveling
perturbations in space-time predicted by
Einstein~\cite{einstein1916naherungsweise, einstein1918gravitationswellen}, and thought to radiate from a variety of
astrophysical processes.
Advanced LIGO recently completed
its first observing period, from
September, 2015 to January, 2016.  
Advanced LIGO is among a
generation of planned instruments that includes
GEO 600, Advanced Virgo, and KAGRA; the capabilities of this
global gravitational-wave network should quickly grow
over the next few years
\cite{Luck:2010rt, TheVirgo:2014hva, Aso:2013eba, 2013arXiv1304.0670L}.

An important class of sources for gravitational-wave detectors are
short duration transients, known collectively as gravitational-wave
bursts \cite{Andersson:2013mrx}.
To search broadly for a
wide range of astrophysical phenomena,
we employ unmodelled searches for gravitational-wave bursts
of durations $\sim$$10^{-3} - 10\,\mathrm{s}$,
with minimal assumptions about the
expected signal waveform.
Bursts may originate from a range of astrophysical
sources, including core-collapse supernovae of massive stars
\cite{Fryer:2011zz} and cosmic string cusps \cite{PhysRevD.71.063510}.
An important source of gravitational-wave transients are 
the mergers of binary black
holes (BBH) \cite{thorne:87, 1989LIGOproposal, Flanagan:1997kp}.  
Burst searches in data from the initial generation of
interferometer detectors were sensitive to distant BBH signals
from mergers with total masses in the range $\sim 20$ -- $400 ~\Msun$
\cite{Aasi:2014iwa, Mohapatra:2014rda}.
Since burst methods do not require
precise waveform models,
the unmodelled search space may include
BBH mergers with mis-aligned spins, large mass ratios, or eccentric orbits.
A number of all-sky, all-time burst searches
have been performed on data from initial LIGO and Virgo
\cite{Abadie:2012rq, Abbott:2015vir, Abadie:2010mt}.
Recent work has focussed on improving detection confidence in unmodelled searches,
and the last year has seen several improvements in the ability to
distinguish astrophysical signals from noise transients
\cite{Thrane:2015psa, Kanner:2015xua, Littenberg:2015kpb, oLIBprep, cwb2g}.
As a result, burst searches
are now able to make high confidence detections across a wide parameter space.

On September 14, 2015, an online burst search~\cite{klimenko:2008fu} reported a transient
that clearly stood above the expected background from detector
noise~\cite{GW150914-DETECTION}.  The alert came only three minutes after the event time-stamp
of \OBSEVENTTIME{} UTC.  A second online burst
search independently identified the event with a latency of a few hours,
providing a rapid confirmation of the signal~\cite{oLIBprep}.
The initial waveform
reconstruction showed a frequency evolution that rises in time, suggesting 
binary coalescence as the likely progenitor, and a best fit model provided
a chirp mass around $28 ~M_{\odot}$, indicating the presence of a BBH signal.  
Within days of the event,
many follow-up investigations began, including detailed checks of the
observatory state to check for any possible anomalies \cite{GW150914-DETCHAR}.
Two days after the signal was found, a notice with the estimated source position
was sent to a consortium of astronomers to search for possible counterparts
\cite{GW150914-EMFOLLOW}.
Investigations continued over the next several months to validate the observation,
estimate its statistical significance, and characterize the astrophysical
source \cite{GW150914-PARAMESTIM, GW150914-CBC}.
 
In this article, we present details of the burst searches that made the 
first detection of the gravitational-wave transient, GW150914 announced
in \cite{GW150914-DETECTION}. We describe results reported in this announcement
that are based on the coherent Waveburst algorithm,
along with those obtained by two other analyses using
omicron-LALInference-Bursts and BayesWave
\cite{klimenko:2008fu, oLIBprep, Cornish:2014kda}.
In Section \ref{data}, we present a brief overview of the
quality of the acquired data and detector performance,
before moving on, in Section \ref{searches}, to present the three analyses employed.
Using each pipeline, we assess the statistical significance of the event.
Section \ref{mdc}  
characterizes each search sensitivity using simulated signals from BBH
mergers.
In Section \ref{pe}, we demonstrate how a range of source properties may be estimated using these
same tools -- including sky position and masses of the black holes.
The reconstructed signal waveform is directly compared to results from numerical
relativity simulations (\nr{}), giving further evidence that this signal is consistent with
expectations from general relativity.  Finally, the paper concludes with a discussion about the implications
of this work.

\section{Data quality and background estimation} \label{data}

\label{data}

We identify 39
calendar days of Advanced LIGO data,
from September 12th to
October 20th, 2015, as a data set to measure the
sensitivity of the searches and the impact of 
background noise events, known as glitches.

As in previous LIGO, Virgo and GEO transient searches
\cite{Abbott:2015vir, Abadie:2012rq, Abadie:2010mt}, 
a range of monitors tracking environmental noise and the state of 
the instruments are used to discard periods of poor quality data.
Numerous studies are performed to identify efficient veto
criteria to remove non-Gaussian noise features, while
having the smallest possible impact on detector livetime \cite{GW150914-DETCHAR}.

However, it is not possible to remove all noise glitches
based on monitors. This leaves a background residual that
has to be estimated from the data.
To calculate the background rate
of noise events arising from
glitches occurring simultaneously at 
the two LIGO sites by chance
\cite{Abadie:2012rq, Abbott:2015vir, Abadie:2010mt},
the analyses are repeated on
$\mathcal{O}(10^6)$ independent time-shifted 
data sets. Those data sets are generated by translating the time of data in one interferometer by a delay of some integer number of seconds, much larger than the maximum GW travel time $\simeq 10$ ms between the Livingston and Hanford facilities.
By considering the whole coincident livetime resulting from
each artificial time shift, we obtain thousands of years of
effective background based on the available data.
With this approach, we estimate a false alarm rate (FAR)
expected from background for each pipeline.

The ``time-shift'' method is effective to estimate the
background due to uncorrelated noise sources at the
two LIGO sites. For the time immediately around GW150914, we also
examined potential sources of correlated noise between
the detectors, and concluded that all possible sources were
too weak to have produced the observed signal \cite{GW150914-DETCHAR}.

\section{Searches for gravitational wave bursts} \label{searches}

Strain data are searched by gravitational wave burst search
algorithms without assuming any particular signal morphology,
origin, direction or time. Burst searches are performed in two
operational modes; on-line and off-line. 

On-line, low-latency searches provide alerts
within minutes of a GW signal passing the detectors to facilitate follow-up
analyses such as searching for electromagnetic counterparts.
In the days and weeks following the data collection, burst analyses
are refined using updated information on the data quality and detector
calibration to perform off-line searches.  
These off-line searches provide improved detection confidence estimates
for GW candidates, measure search sensitivity, and add to waveform reconstruction and
astrophysical interpretation.
For short-duration, narrowband signals, coherent burst searches
have sensitivities approaching that of optimal matched filters 
\cite{Anderson:2000yy, Mohapatra:2014rda}.

In the following subsections, we describe the burst analysis of GW150914. This includes two independent end-to-end pipelines, coherent Waveburst (cWB)
and omicron-LALInference-Bursts (oLIB), and BayesWave, which performed a follow-up analysis at trigger times identified by cWB.
These three algorithms employ different strategies (and implementations) to search for unmodelled GW transients, hence, they could perform quite differently for specific classes of GW signals. Given the very broad character of burst signals, the use of multiple search algorithms is then beneficial, both to validate results and to improve coverage of the wide signal parameter space.

A summary of the results from cWB has been presented in~\cite{GW150914-DETECTION}. Here, we provide more details regarding the cWB search pertaining the discovery of GW150914 and present its results with respect to the other burst searches.
In this paper, we focus our characterizations of our pipelines on
BBH sources only.

\subsection{Coherent WaveBurst}

The cWB algorithm has been used to perform all-sky searches for gravitational
wave transients in LIGO, Virgo and GEO data since 2004. The most recent cWB 
results from the initial detectors are ~\cite{Abadie:2010mt,Abadie:2012rq,IMBHB:S6}.  
The cWB algorithm has since been upgraded to conduct transient searches with the
advanced detectors~\cite{cwb2g}.  The cWB pipeline was used in the low-latency
transient search that initially detected GW150914, reporting the event three
minutes after the data was collected. This search aims at rapid alerts for the
LIGO/Virgo electromagnetic followup program~\cite{GW150914-EMFOLLOW} and provides a
first estimation of the event parameters and sky location.  A slightly different
configuration of the same pipeline was used in the offline search to measure the
statistical significance of the GW150914 event which is reported
in~\cite{GW150914-DETECTION}.  The low-latency search was performed in the frequency range
of 16-2048~Hz, while the offline search covered the band of the best detector
sensitivity between 16 and 1024~Hz.

\subsubsection{cWB pipeline overview} \label{cwb_overview} The cWB pipeline
searches for a broad range of gravitational wave transients in the LIGO
frequency band without prior knowledge of the signal
waveforms~\cite{klimenko:2008fu}. The pipeline identifies coincident events in data from the two LIGO detectors 
and reconstructs the gravitational-wave signal
associated with these events using a likelihood analysis.

First, the data are whitened and converted to the
time-frequency domain using the Wilson-Daubechies-Meyer wavelet
transform~\cite{Necula:2012}. Data from both detectors are then combined 
to obtain a time-frequency power map. A transient event is identified as a cluster of
time-frequency data samples with power above the baseline detector noise. To obtain a
good time-frequency coverage for a broad range of signal morphologies, the
analysis is repeated with seven frequency resolutions $\Delta{f}$ ranging from
1~Hz to 64~Hz in steps of powers of two, corresponding to time resolutions
$\Delta{t}=1/(2\Delta{f})$ from $500$~ms to $7.8$~ms. The clusters at different
resolutions overlapping in time and frequency are combined into a trigger that
provides a multi-resolution representation of the excess power event recorded by the detectors.

The data associated with each trigger are analyzed
coherently~\cite{cwb2g} to estimate the signal waveforms, the wave polarization, and
the source sky location. The signal waveforms in both detectors are 
reconstructed with the constrained likelihood method~\cite{klimenko05}. 
The constraint used in this analysis is model independent and requires the 
reconstructed waveforms to be similar in both detectors, as expected from the close 
alignment of the H1 and L1 detector arms.

The waveforms are reconstructed over a uniform grid of sky locations with
$0.4^\circ \times 0.4^\circ$ resolution.  We select the
best fit waveforms that correspond to the maximum of the likelihood statistic
$L=c_cE_s$, where $E_s$ is the total energy of the reconstructed 
waveforms\footnote{$\sqrt{E_s}$ is the network signal-to-noise ratio \cite{cwb2g}} 
and $c_c$ measures the similarity of the waveforms in the two detectors. 
The coefficient $c_c$ is defined as $c_c=E_c/(E_c+E_n)$, where $E_c$ is 
the normalized coherent energy and $E_n$
is the normalized energy of the residual noise after the reconstructed signal is
subtracted from the data. The coherent energy $E_c$ is proportional to the
cross-correlation between the reconstructed signal waveforms in H1 and L1 detectors.
Typically, gravitational wave signals are coherent and have small residual energy i.e., $E_c \gg
E_n$ and therefore $c_c \sim 1$. On the other hand, spurious noise events (glitches) 
are often not coherent, and have large residual energy because
the reconstructed waveforms do not fit well the data
i.e., $E_c \ll E_n$ and therefore $c_c \ll 1$.
The ranking statistic is defined as $\eta_c=(2c_cE_c)^{1/2}$. By construction, 
it favors gravitational-wave signals correlated in both detectors 
and suppresses un-correlated glitches.

\subsubsection{Classification of cWB events}

Events produced by the cWB pipeline with $c_c > 0.7$ are selected and divided
into three search classes $C1$, $C2$, and $C3$
according to their time-frequency morphology. The purpose of this
event classification is to account for the non-Gaussian noise that
occurs non-uniformly across the parameter space searched by the pipeline.

The classes are determined by three algorithmic tests and additional selection
cuts. The first algorithmic test addresses a specific type of noise transient
referred to as ``blip glitches''~\cite{GW150914-DETCHAR}. During the run, both detectors
experienced noise transients of unknown origin consisting of a few cycles around
100~Hz.
These blip glitches have a very characteristic time-symmetric waveform
with no clear frequency evolution.
Previous work has shown that down-weighting signals with simple time-frequency
structure can enhance pipeline performance \cite{Kanner:2015xua}.
To implement this here, we apply a
test that uses waveform properties to identify,
in the time domain, blip glitches occurring at both detectors. The second
algorithmic test identifies glitches due to non-stationary narrow-band
features, such as power and mechanical resonance lines. This test
selects candidates which have most of their energy
(greater than 80\%) localized in a frequency bandwidth  less than 5~Hz.  A cWB
event is placed in the search class $C1$, if it passes either of the
aforementioned tests. In addition, due to the elevated non-stationary 
noise around and below the Advanced LIGO mechanical resonances at 41 Hz,  
events with central frequency lower than 48 Hz were 
also placed in the $C1$ class.

The third algorithmic test is used to identify events with a frequency increasing
with time. The reconstructed time-frequency patterns can be characterized by an
ad-hoc parameter $\mathcal{M}$ following Eq.~(\ref{inteqchirpmassfreq}) in
Sec.~\ref{chirpmass}. For coalescing binary signals $\mathcal{M}$
corresponds to the chirp mass of the binary \cite{PhysRev.136.B1224}. 
For signals that do not originate from coalescing binaries and glitches, $\mathcal{M}$ takes on unphysical values.
In the un-modeled cWB analysis, the parameter $\mathcal{M}$ is used to 
distinguish between events with different time-frequency evolution. 
By selecting events with $\mathcal{M}>1 \Msun$ we
identify a broad class of events with a chirping time-frequency signature, which 
includes a subclass of coalescing binary signals.  The
events selected by this test that also have a residual energy $E_n$ consistent with Gaussian
noise are placed in the search class $C3$.  All other events, not
included into the $C1$ or $C3$ class, are placed in the search class $C2$. 
The union of all three independent search classes covers the full
parameter space accessible to the unmodelled cWB search.

\subsubsection{False alarm rate}

\begin{figure}
\mbox{
\includegraphics[width=\columnwidth]{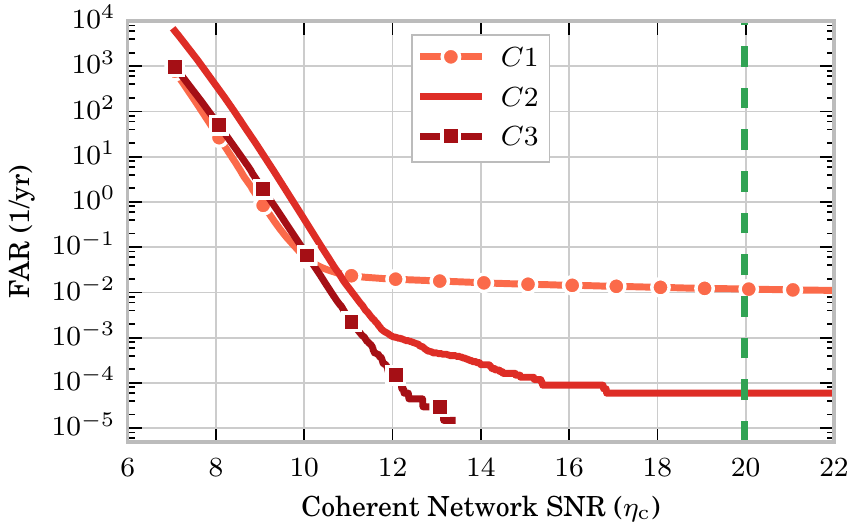} 
}
\caption{Cumulative rate distribution of background events as a function of the
detection statistic $\eta_c$ for the three cWB search classes. 
Vertical dashed line shows the value of the detection statistic for the GW150914 event. 
}
\label{fig:cWB_bkg}
\end{figure}

To establish the distribution of background  events, we use the time-shift procedure
discussed in Sec. \ref{data}, using all the data available for each detector.
The effective background livetime for this analysis is $\BURSTBCKLIVETIME$~years, 
obtained by analysing more than $1.6 \times 10^6$ time-shifted instances of 16 days of 
the observation time.
Figure~\ref{fig:cWB_bkg} reports the cumulative false alarm rate distributions 
as a function of the detection statistic $\eta_c$ for the three defined search classes.
The significance of a candidate event is measured against the background of its class.
As shown in the plot, the $C1$ search class is affected by a tail of blip glitches 
with the false alarm rate of approximately $0.01~y^{-1}$.
Confining glitches in the $C1$ class enhances the search
sensitivity to gravitational-wave signals falling in the $C2$ and $C3$
classes. In fact, the tail is reduced by more than two orders of magnitude in the $C2$ search class. 
The background rates in the $C3$ search class are almost ten times lower
than in 
$C2$, with no prominent tail of loud events, indicating that it is highly 
unlikely for detectors to produce coherent background events with 
a chirping time-frequency evolution.

To check the homogeneity and stability of background rates shown in Figure
\ref{fig:cWB_bkg}, these distributions have been compared between instances of
background data, generated with different time-shifts between the detectors,
finding no evidence for any dependence on the time-shift interval or on the
time-period of data collection.

\subsubsection{Significance of GW150914 event}

\begin{figure*}
\mbox{
\includegraphics[width=\columnwidth]{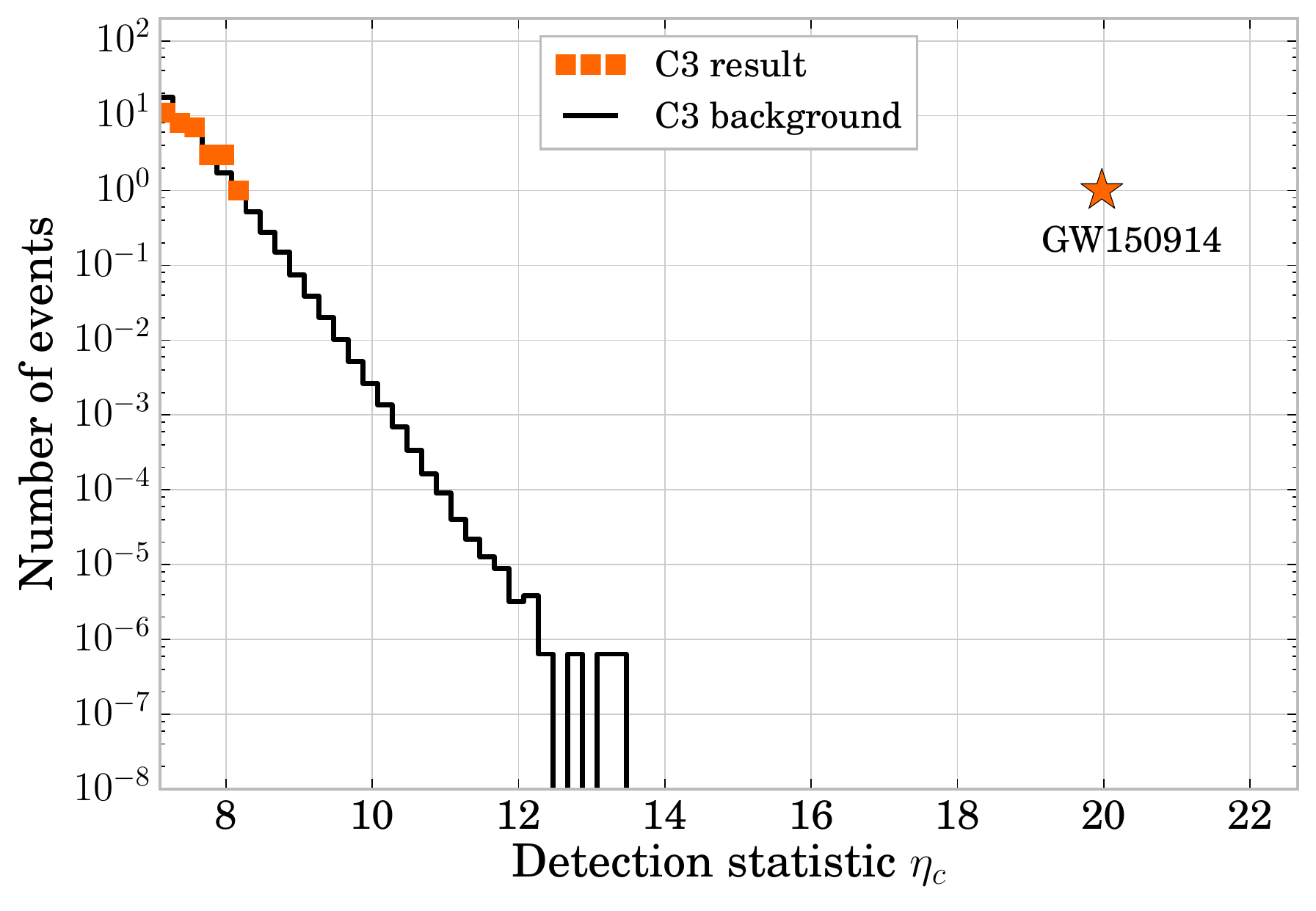} 
\includegraphics[width=\columnwidth]{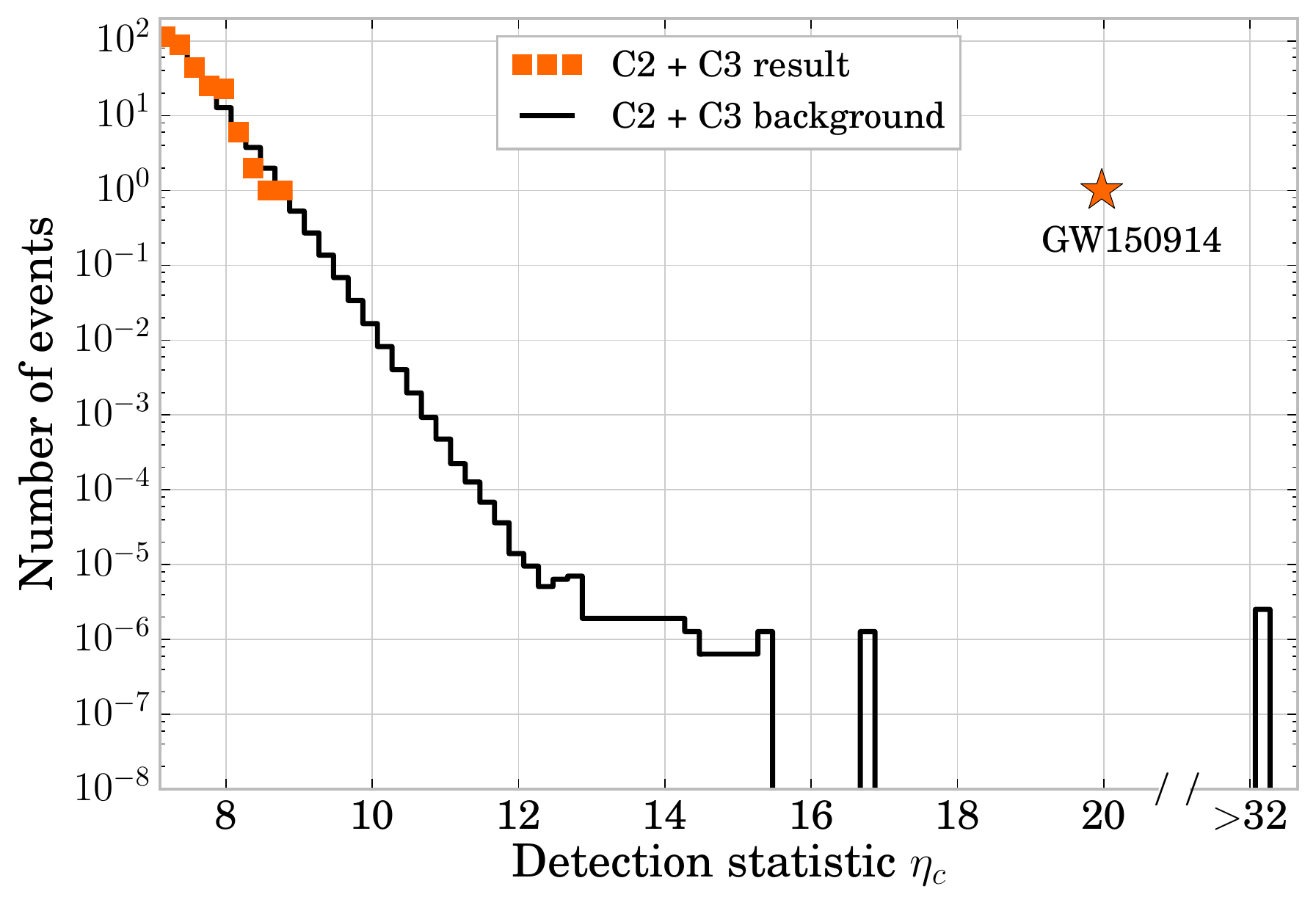} 
}
\caption{Search results (in orange) and expected number of background events (black) in 16 day 
of the observation time as a function of the cWB detection statistic
(bin size 0.2)
for the $C3$ search class (left) and $C2+C3$ search class (right).
The black curve shows the total number of background events found in \BURSTBCKLIVETIME ~years of data, rescaled to 16 days of observation time.
The orange star represents GW150914, found in the $C3$ search class. 
}
\label{fig:cWB_C3bin}
\end{figure*}

GW150914 was detected with $\eta_c=20$ and belongs to the $C3$ class.  Its
$\eta_c$ value is larger than the detection statistic of all observed cWB candidates.
Also Figure~\ref{fig:cWB_C3bin} (left) shows that the GW150914 $\eta_c$ value
is larger than the detection statistic of any
background event in its search class in \BURSTBCKLIVETIME ~years of
the equivalent observation time. All other observed event candidates
(orange squares) are consistent with the background.

The GW150914 significance is defined by its false alarm rate measured 
against the background in the $C3$ class.
Assuming that all search classes are statistically independent, 
this false alarm rate should be increased by a conservative trials factor 
equal to the number of classes. By taking into account the trials factor of 3, 
the  estimated GW150914 false alarm rate is less than one event in \BURSTEVENTIFAR ~years.
The probability that the 16 days of data would yield a noise event with this false alarm rate is less than  
$16 / (365 \times \BURSTEVENTIFAR) = 2 \times 10^{-6}$.

The union of the $C2$ and $C3$ search classes represents a transient search with no 
assumptions on the signal time-frequency evolution. The result of such analysis
with just two search classes $C1$ and $C2+C3$ is shown in Figure~\ref{fig:cWB_C3bin} (right).
In this case there are four events louder than GW150914 in the $C2+C3$ class. 
With the trials factor of 2, the false alarm rate is one event in 8\,400 years.
The four loud events are produced by a random coincidence of multiple blip glitches: 
two nearby blip glitches in one detector and a single blip glitch in the second detector. 
The algorithmic test that identifies blip glitches
was not designed to capture multiple ones and, therefore, missed these events.

\subsection{oLIB}

The oLIB search~\cite{oLIBprep} is a search pipeline for gravitational-wave
bursts designed to operate in low-latency, with results typically produced in
around thirty minutes. However, the pipeline can operate in two modes, online
and offline. The online version identified GW150914 independently of cWB. The
offline version is used here to establish the significance of GW150914.

\subsubsection{oLIB pipeline overview}

The oLIB pipeline follows a hierarchical scheme first performing a coincident
event down-selection followed by a fully-coherent Markov chain Monte Carlo
Bayesian analysis.

In the first step of the pipeline, a time-frequency map of the
single-interferometer strain data from all detectors is produced using the
Q-transform \cite{chatterji-2004-21} implemented in
Omicron~\cite{OmicronRobinet}. Stretches of excess power are flagged as
triggers. Neighboring triggers that occur within 100~ms, with an identical
central frequency $f_0$ and quality factor $Q$ are clustered together. After
applying data quality vetoes as described in Sec. \ref{data}, a list of triggers
that fall within a 10~ms coincidence window (compatible with the speed-of-light
baseline separation of the detectors) is then compiled.

In the second step of the pipeline, all coincident triggers identified in the
first step are analyzed using LIB, a Bayesian parameter estimation and model
selection algorithm that coherently explores the signal parameter space with the
nested sampling algorithm~\cite{Skilling} available in the LALInference software
library~\cite{Veitch:2014wba}.

LIB models signals and glitches by a single sine-Gaussian wavelet.  Signals have
a coherent phase across detectors, while glitches have not. Using this model,
LIB calculates two Bayes factors, each of which represents an evidence ratio
between two hypotheses: coherent signal vs Gaussian noise (BSN) and coherent
signal vs incoherent glitch (BCI). These two Bayes factors are then combined
into a scalar likelihood ratio $\Lambda$ for the signal vs noise (Gaussian or
glitch) problem. More precisely, $\Lambda$ is obtained from the ratio of the
probability distributions for the Bayes factors BSN and BCI estimated
empirically from ``training'' sets of events. Those sets consists in $\simeq 4000$
simulated gravitational wave signals from a uniform-in-volume source
distribution and $\simeq 150$ background triggers obtained from time-shifted
data for the signal and noise cases, respectively.

The final ranking statistic $\Lambda$ is evaluated for a different set of
background triggers from time-shifted data in order to map a given value of the
likelihood ratio into a FAR.

\subsubsection{oLIB analysis of GW150914}

For the purpose of this analysis, Omicron runs over the $32 - 1024 \mathrm{Hz}$
bandwidth and selected triggers that exceed a SNR threshold of 6.5. LIB uses
the following priors: uniform in sky location, uniform in central frequency
$f_0$ in the selected bandwidth, and uniform in quality factor $Q$ from $0.1 -
110$. Events with $\text{BSN}$ or $\text{BCI} \leq 0$ are discarded. We retain
events with $48 \leq \tilde{f}_{0} \leq 1020 \mathrm{Hz}$ and
$2 \leq \tilde{Q} \leq 109$
where $\tilde{f}_{0}$ and $\tilde{Q}$ are median values computed from
the posterior distributions delivered by LIB. The selection cut on $\tilde{Q}$
is analogous to those used by cWB to reject blip glitches and narrow-band
features.  The ranking statistic $\Lambda$ and its background distribution from
which the FAR is deduced are computed from the training and background sets
after applying all those cuts.

\begin{figure}
\mbox{
\includegraphics[width=\columnwidth]{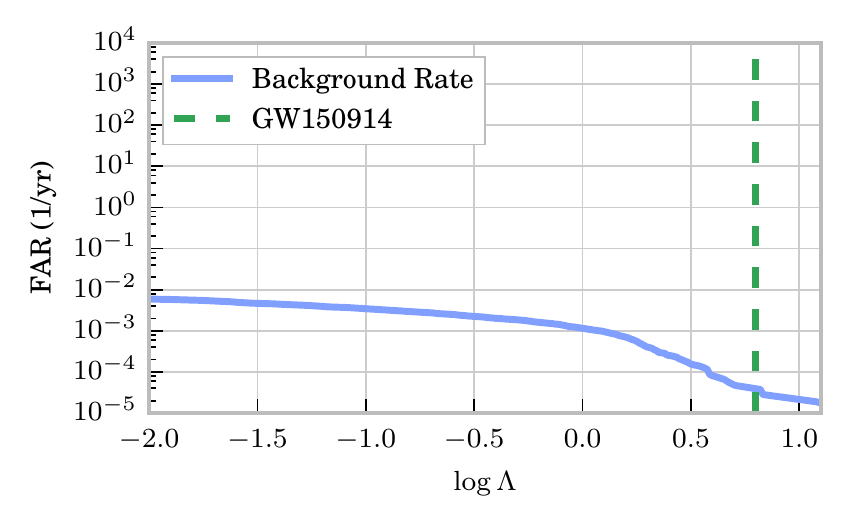} 
}
\caption{Cumulative rate distribution of background events as a function of oLIB ranking statistic $\log{\Lambda}$.
GW150914 is the only event in the search results to pass all thresholds. Its statistic value $\log{\Lambda} = 0.80$ corresponds to a background FAR of $\simeq$ 1 in 27\,000 years. 
}
\label{fig:olib_bg}
\end{figure}

Because oLIB is able to run on short data segments ($\gtrsim 3$ s), this search
analyzed nearly all available data, which amounted to 17.4 days, i.e, $\sim
10\%$ more coincident data than cWB.  The data were time-shifted in one-second
intervals to produce the equivalent of 106\,000 years of background data. The
background distribution is plotted as a function of $\log{\Lambda}$ in Figure
\ref{fig:olib_bg}. As shown in the same figure GW150914 has a ranking statistic
of $\log{\Lambda} = 0.80$, corresponding to a FAR of roughly 1 in 27\,000
years. It is the only event in the search results satisfying the selection cuts.

\subsection{BayesWave follow-up}

\begin{figure}
\mbox{
\includegraphics[width=\columnwidth]{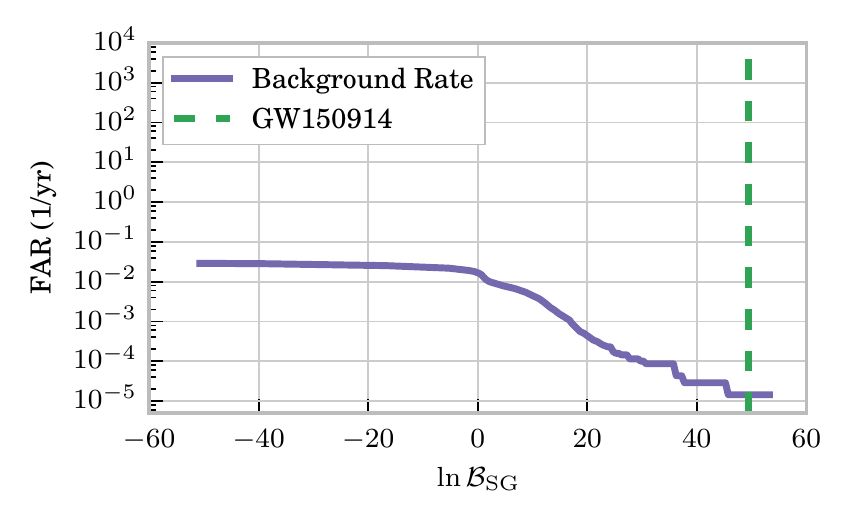} 
}
\caption{Cumulative rate distribution of background events as a function of the cWB+BayesWave detection statistic $\ln \mathcal{B}_{SG}$.
The cWB+BayesWave pipeline considers all cWB candidates with $\eta_c>11.3$ (combining all three curves in Fig.~\ref{fig:cWB_bkg}).
In the equivalent of \BURSTBCKLIVETIME ~years of data, GW150914 was the only zero-lag event to pass all thresholds. Only one noise coincidence is ranked higher than GW150914. 
}
\label{fig:bw_bg}
\end{figure}

The BayesWave pipeline is a Bayesian algorithm designed to robustly distinguish GW signals from glitches in the detectors~\cite{Cornish:2014kda,Littenberg:2014oda}.
In this search, BayesWave is run as a follow-up analysis to
triggers identified by cWB.
For each candidate event, BayesWave compares the marginalized likelihood, or evidence, between three hypotheses: The data contain only
Gaussian noise; the data contain Gaussian noise and noise transients (glitches); or the data contain Gaussian
noise and an astrophysical signal.

The BayesWave algorithm models signals and glitches using a linear combination of sine-Gaussian wavelets.
The number of wavelets needed in the glitch or signal model is not fixed \textit{a priori},
but instead is optimized using a reversible jump Markov chain Monte Carlo. 
The glitch model fits the data separately in each interferometer with an independent linear combination of wavelets.
The signal model reconstructs the candidate event at some fiducial location (the center of the Earth), taking into account the response of each detector in the network to that signal.
BayesWave uses a parameterized phenomenological model, BayesLine, for the instrument noise spectrum, simultaneously characterizing the Gaussian noise and instrument/astrophysical transients~\cite{Littenberg:2014oda}.

BayesWave produces posterior distributions for the parameters of each model under
consideration.
For the signal model, this includes the waveform, as constructed from sums 
of sine-Gaussian wavelets, and the source position.
Waveform reconstructions are used to produce posterior distributions for
characteristics such as the duration, central frequency, and bandwidth of the signal,
which are used to compare the data to theoretical models.  The marginalized
posterior (evidence) for each model is calculated
by marginalizing over the different dimension waveform reconstructions,
and then is used
to rank the competing hypotheses.

BayesWave is used as a follow-up analysis for candidate events first identified by cWB.
The combined cWB+BayesWave data analysis pipeline has been shown to
allow high-confidence detections across a range of waveform
morphologies \cite{Kanner:2015xua, Littenberg:2015kpb}.
The cWB+BayesWave pipeline uses the Bayes factor' comparing the signal and glitch models 
($\mathcal{B}_{SG}$) as its detection statistic.
Bayes factors are reported on a natural logarithmic scale $\ln\mathcal{B}_{SG}$, which scales
with $N\ln{\mathrm{SNR}} $,
where $N$ is the number of wavelets used in the reconstruction \cite{Littenberg:2015kpb}.
The consequence is that BayesWave assigns a higher detection statistic to
signals with non-trival time-frequency structure.
Though Bayes factors used by Bayeswave and oLIB methods both produce a measure of coherence between the signal morphologies
observed in multiple detectors, the above calculation indicates that BayesWave, $\mathcal{B}_{SG}$ also includes a measure of  the signal complexity.
	
The ``offline'' BayesWave pipeline analyzes all cWB zero-lag and background events with a
detection statistic $\eta_c > 11.3$ and correlation coefficient $c_c>0.7$.
The threshold on $\eta$ for event follow-up is a compromise between computational cost and in-depth analysis of cWB events.
The BayesWave computation is performed over a four-second segment of data\footnote{The four second segment length was shown in testing to be the minimum amount of data needed to estimate the PSD.} centered on the event time reported by cWB.
We use one second of data around the event time for model comparison, while the remainder of the segment is used for spectral estimation.
We perform the analysis in the Fourier domain over the frequency range of $32 < f < 1024$ Hz though, for cWB candidates with central frequency $f_{\rm cWB} < 200$ Hz (including GW150914), BayesWave used a maximum frequency of 512 Hz to reduce the computational cost of the analysis.
Both the signal and the glitch model require at least one wavelet (to make them disjoint from one another and the Gaussian noise model) and have a maximum of 20 wavelets allowed in the linear combination.
Most of the priors used in the analysis are as described in \cite{Cornish:2014kda} and \cite{Littenberg:2015kpb},
with the following changes.
The prior on the ``quality factor'' of the wavelets $Q$ has been extended to include lower
values, so that it is uniform
over the interval [0.1,40].
The low $Q$ values
allow blip glitches to be correctly characterized with a small number of wavelets.
Also, the functional form of the glitch amplitude prior has been modified to scale
as a power law rather than an exponential in the large SNR limit.  The new prior better
reflects the belief that very loud events (${\rm SNR} > 100$) are more likely to be
glitches than signals.

Figure~\ref{fig:bw_bg} shows the cumulative rate distribution of background events as a function of the cWB+BayesWave detection statistic $\ln \mathcal{B}_{SG}$.
The cWB+BayesWave pipeline considers the triggers from all cWB search classes together 
(all curves in Fig.~\ref{fig:cWB_bkg}) as a single search.
The explicit glitch model used by BayesWave
reduces the tail in the background distribution ~\cite{Littenberg:2015kpb},
so that loud background events are down-weighted rather than grouped into different classes.
In the equivalent of \BURSTBCKLIVETIME ~years of O1 data,
2374 cWB events warranted a BayesWave follow up and only one noise coincidence ($ \ln \mathcal{B}_{SG} = 53.1 \pm 3.4 $) was ranked higher than GW150914
($ \ln \mathcal{B}_{SG} = 49.4 \pm 0.8$).  GW150914 is the only zero-lag event to pass all thresholds.
Investigations of the highest ranking background events have revealed remarkably similar glitches in the two detectors which, were it not for the large, unphysical time shifts applied to the data, would be indistinguishable from a GW signal.
However, the waveform morphology of the most significant background events is in no way similar to a BBH merger signal.  Treating all cWB candidates as
coming from the same search, BayesWave estimates a FAR for GW150914 of
1 in \BURSTBCKLIVETIME ~years.

\section{Search Sensitivity} \label{mdc}

In this section, we demonstrate the ability of transient searches to detect GWs from BBH mergers.
We use simulated gravitational waveforms that cover all three phases of BBH
coalescence, i.e., inspiral, merger and ringdown. The analysis is performed by adding
simulated BBH waveforms to the detector data, and recovering them using the three burst pipelines
described in Sec.~\ref{searches}.

\subsection{Simulation data set}

BBH systems are characterized by the masses $m_1$ and $m_2$, and
dimensionless spin vectors $\mathbf{a}_1$ and $\mathbf{a}_2$ of the two
component black holes, the source distance $D$, its sky-location coordinates,
and the inclination of the BBH orbital momentum vector relative to the
line-of-sight to Earth. The black-hole spins are obtained from the
dimensionless spin vectors by $\mathbf{S}_i = m_i^2 \mathbf{a_i}$ where
$|\mathbf{a}_i| \leq 1$.

The simulation includes binaries that are isotropically located on the sky and
isotropically oriented, with total masses $M=m_1+m_2$ uniformly distributed
between 30 and 150 $\Msun$, that is within a factor of $\sim 2$ of the estimated
total mass for GW150914 \cite{GW150914-PARAMESTIM}.  We generate three separate sets, each with a
fixed mass-ratio $q=m_2/m_1 \in \left\{0.25, 0.5, 1.0\right\}$. We assume that
the black-hole spins are aligned with the binary orbital angular momentum, with
a spin magnitude uniformly distributed across $|\mathbf{a}_{1,2}| \in
[0,\,0.99]$.
The distances are drawn from a distributions within $3.4$ Gpc such that we get
good sampling for a range of SNR values around the detection threshold.
The simulation does not include red-shift corrections, which introduces small
systematic errors for the more distant sources.
The signals are distributed uniformly in time with a gap of 100 seconds between them.

\begin{figure*}[!htb]
  \includegraphics{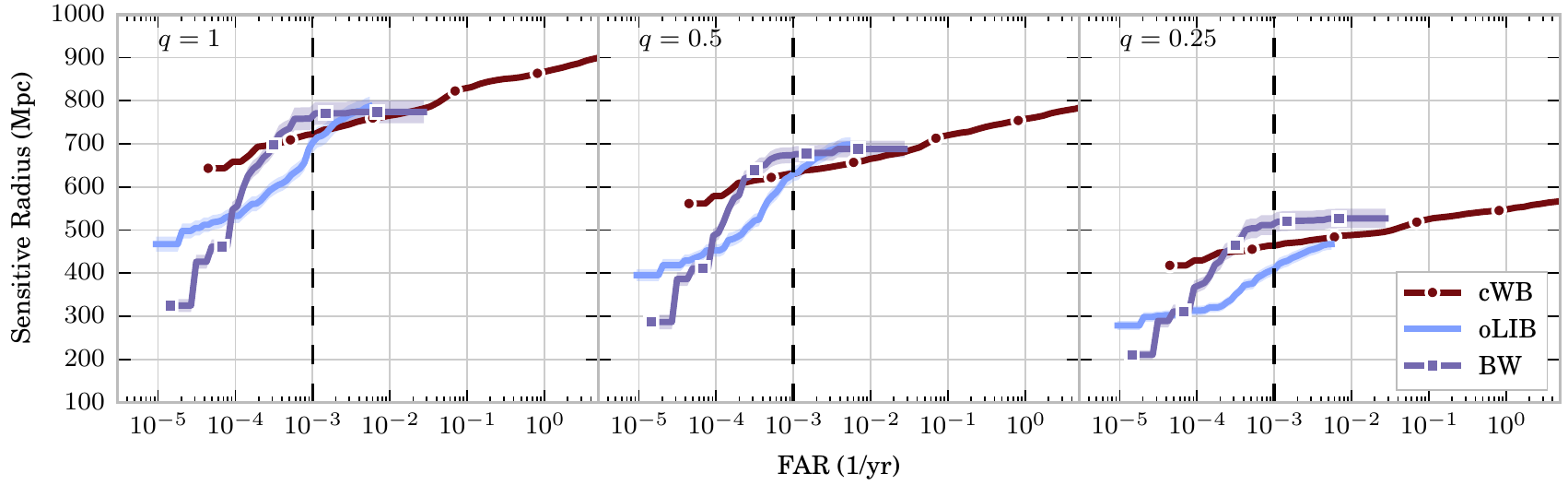}
  \caption{Sensitive radius for the different search pipelines for simulated BBH
    waveforms with different mass ratios $q$. The sensitive radius measures the
    average distance to which the search detects with a given FAR threshold.  The cWB results include all three search classes, with a corresponding trials factor.}
\label{f:cwb-pycbc}
\end{figure*}

The BBH waveforms analyzed in this study have been generated using the \texttt{SEOBNRv2} model
in the \texttt{LAL} software library \cite{Taracchini:2013rva, Kumar:2015tha}.
This model only accounts for the dominant $\ell=2, m=2$ GW radiated modes. The
waveforms are generated with an initial frequency of $15$ Hz. The data sets are
summarized in Table \ref{tbl:inj}.

\begin{table}[h]
  \begin{center}
  \begin{tabular}{l c r }
   		\hline 
  		\hline 
    Total mass $M=m_1+m_2$ & & $30-150 M_\odot$  \\
    Mass matio, $q = m_2/m_1$ & & 0.25, 0.5, 1.0  \\
    Spin magnitude $|\mathbf{a}_{1,2}|$ & & 0 -- 0.99  \\
    Waveform model & & \texttt{SEOBNRv2}\\
      		\hline 
  	\end{tabular}
  \end{center}
  \caption{ Summary of the BBH simulations used for estimating search efficiency.  }
  \label{tbl:inj}
\end{table}

\subsection{Results}
To quantify the results of the study, we use the sensitive radius which is the radius of the sphere with volume $V=\int 4\pi r^2 \epsilon(r) dr$, where $\epsilon(r)$ is the averaged search efficiency for sources at distance $r$ with random sky position and orientation \cite{Mohapatra:2014rda}. 
 For each pipeline,
we calculate the sensitive radius as a function of FAR.  The results are shown in Figure
\ref{f:cwb-pycbc}.
For example, at a FAR of 1 per thousand years, the three searches show
similar performance, with each
detecting the simulated equal-mass BBH population to a sensitive distance in the range
700 to 800 Mpc.  To the far left side of the plots (very low FAR),
the differences between pipelines are dominated by the loudest few background
events; the cWB $C3$ search class selection for chirping events allows many
BBH signals to be recovered with very low FAR.

\begin{figure}[]
  \includegraphics{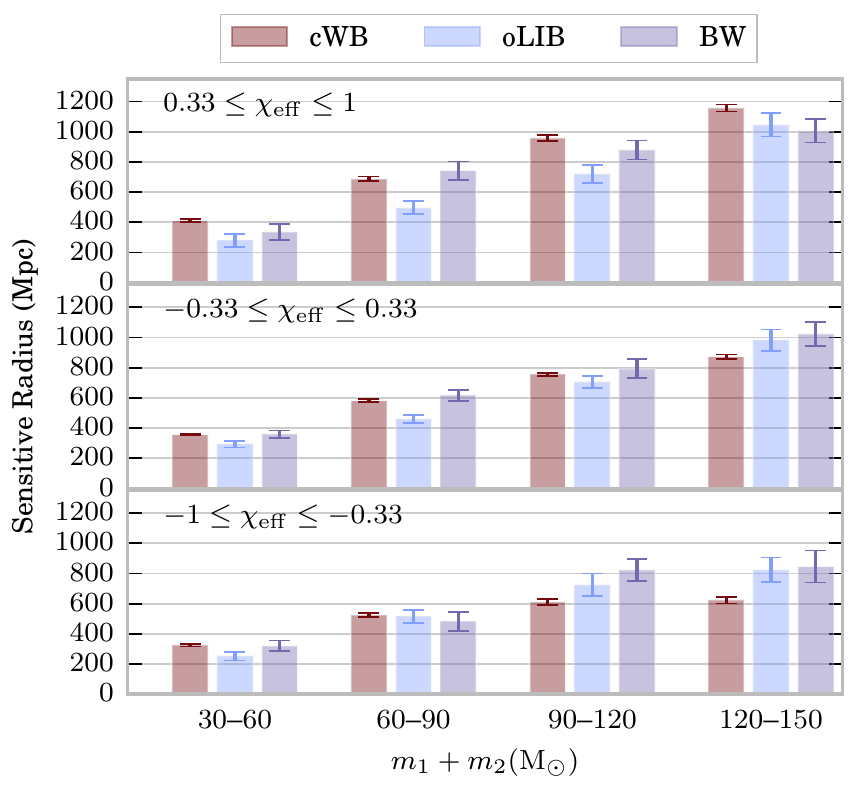}
  \caption{{\bf Dependence of sensitive radius on spins of BBH:} To investigate the effect of spins of black holes on the detection of BBH systems, we show the search radius $R$ for each pipeline for varying effective spins with mass-ratio $q = 1$ at FAR = $10^{-3}$ 1/yr.  The total mass range is varied from $30-150 ~\Msun$, while the effective spin is distributed into three bins: aligned spins (${\bf \chi_{eff}} \in [0.33,1]$), anti-aligned (${\bf \chi_{eff}} \in [-1, -0.33]$) and non-spinning (${\bf \chi_{eff}} \in [-0.33, 0.33]$).  The error bars represent the statistical uncertainty of the sample.  The cWB results include all three search classes, with a corresponding trials factor.}
\label{f:sensitivity-spin}
\end{figure}

The effect of intrinsic BBH parameters (component masses and spins)
on the sensitive radius of
the three pipelines is summarized in Figure \ref{f:sensitivity-spin}.
The three panels of the figure correspond to three bins of effective spin.
Effective spin is defined as in~\cite{GW150914-PARAMESTIM}:
 ${\bf \chi_{eff}} = \left( \frac{\bf S_1}{m_1} + \frac{\bf S_2}{m_2} \right) \cdot \frac{\bf \hat{L}}{M} $, with
 ${\bf \hat{L}}$ the direction of orbital angular momentum.
Depending on the mass and spin of the binary, the sensitive radius can vary from about 250\,Mpc up
to over 1\,Gpc.  Over this range, larger masses are detectable to further distances.  Spins which
are aligned with the orbital angular momentum tend to increase the sensitive radius,
while anti-aligned spins make the systems more difficult to detect.  For the
mass/spin bin most like GW150914 - 60-90 $\Msun$,
the sensitive radius of the searches are between $400$ and $600$ Mpc.

\section{Source Characterization} \label{pe}

In \cite{GW150914-PARAMESTIM}, we present estimates for the
parameters of the binary black-hole model that best describes GW150914.
These parameters include the masses and spins of the binary components,
and their posterior distributions represent our most complete description of
the astrophysical source. In this work, we take a complementary approach, by
using the outputs of the burst pipelines described in Section \ref{searches}
to characterize the event.  Many of the burst pipeline outputs are
available in low-latency, so this approach
can inform follow-up studies in a timely fashion. For example, the cWB estimate of the
GW150914 chirp mass was available within minutes, and provided the first
evidence that this signal originated from merging black holes.
Likewise, low-latency position estimates are used for counterpart searches
\cite{GW150914-EMFOLLOW}.  

Burst analyses are also able to estimate the time evolution of observed waveforms,
a process we refer to as waveform reconstruction.
Burst waveform reconstruction algorithms do not rely on astrophysical
models. Instead, estimates of the coherent gravitational-wave power observed
by the detector network are used to reconstruct the signal.  These waveform
reconstructions are valuable: they provide an unbiased view of the signal
most consistent with the observatory data.  Such reconstructed signals
can be used to classify the source type, compare with models, and potentially
identify unexpected features.  In this section, we present how the
outputs of the burst piplines were used to estimate the source position,
reconstruct the waveform, and characterize the BBH source.
We also compare the reconstructed waveforms with a set of numerical relativity
waveforms, in order to check the consistency of our results against the
most precise class of models available.

\subsection{Source localization}\label{s:skymaps}

Three burst algorithms (cWB, BayesWave, and LIB) produce localization estimates for the
GW event.
These ``skymaps'' can be interpreted as the posterior probability distribution
of the source's right-ascension ($\alpha$) and declination ($\delta$) given
the observed data.
cWB produces skymaps during its detection process by maximizing a
constrained-likelihood on a grid over the sky; these
are available within minutes of the candidate's detection.
LIB and BayesWave perform more computationally expensive
analyses, and so produce results with higher latency.
LIB uses a space of single sine-Gaussian waveforms as its waveform model,
and produces skymaps after one to two hours, whereas BayesWave
maps can take as long as several days to be produced, since it explores a larger parameter
space of superpositions of sine-Gaussian waveforms.
Each algorithm makes different and somewhat complementary assumptions about the signal,
and these assumptions affect their localization estimates. By localizing signals with
multiple algorithms, we can cross-check and validate the localization estimate and
identify any systematic difference between the algorithms \cite{Essick:2014wwa}.

An overview of the skymaps
used by astronomers to search for counterparts to GW150914
may be found in \cite{GW150914-EMFOLLOW}, including the
cWB and LIB skymaps.
Here, we compare cWB, LIB, and BayesWave skymaps in addition to the 
map produced by LALInference with binary coalescence templates, which samples the posterior
distribution of all signal parameters using signal waveforms
that cover the inspiral, merger and ringdown phase~\cite{Veitch:2014wba}.
For GW150914, we expect the LALInference map to yield a
relatively precise localization,
because it assumes a waveform from a compact binary coalescence,
instead of the broad waveform classes used by the burst pipelines.
Burst localization algorithms produce systematically larger skymaps than
template-based algorithms because they make fewer assumptions about the waveform.
However, the LALInference map reported here also includes the effects
of calibration uncertainty within the detectors,
which significantly widen the uncertainty of this reconstruction
\cite{GW150914-CALIBRATION}.
In principle, calibration effects could also be included in the
burst skymaps, but what is shown here represents the information
that was available at the time electromagnetic astronomy observations began \cite{GW150914-EMFOLLOW}.

\begin{figure*}
    \includegraphics[width=\textwidth]{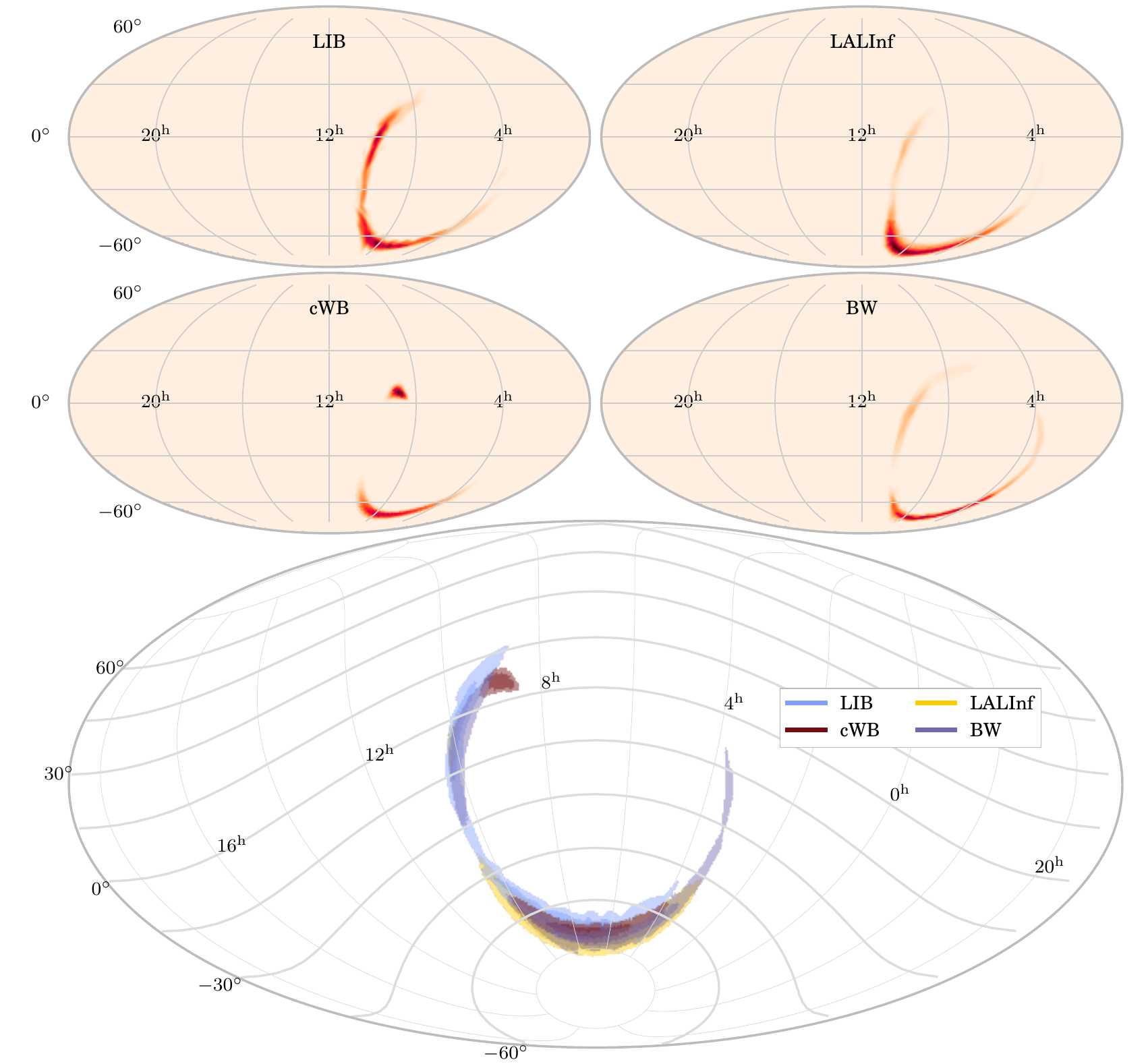}
    \caption{All-sky projections of several skymaps produced for GW150914.
    Above, each map is shown by itself in celestial coordinates. Below, a rotated coordinate system shows contours defining the 50\% and 90\% confidence regions for four reconstructions. }
    \label{f:mollweide corner}
\end{figure*}

\begin{figure}
\mbox{
\includegraphics[width=\columnwidth]{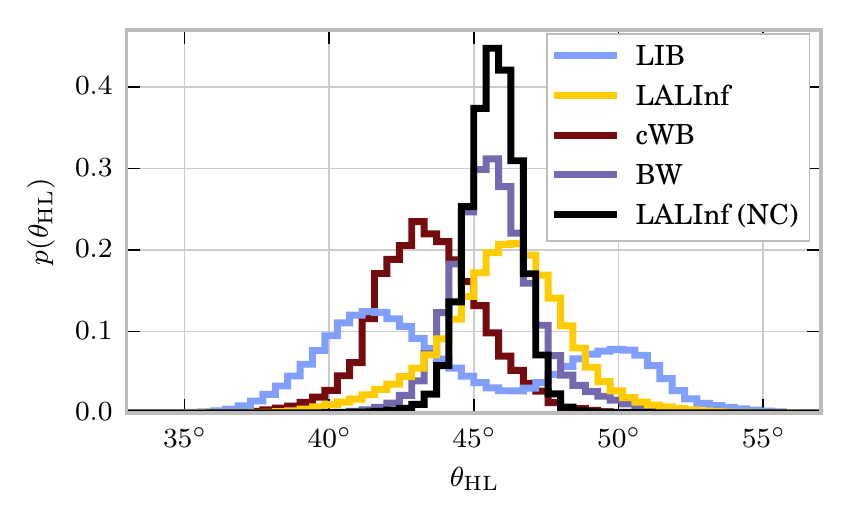}
}
    \caption{Marginal distributions of the polar angle defined by triangulation. These give a measure of the width of each ring.}
    \label{fig:sky_hist}
\end{figure}

Figure \ref{f:mollweide corner} shows Mollweide projections in ($\alpha$,$\delta$)
of all skymaps considered, as well as overlays of the 50\% and 90\% contours in
a rotated frame of reference.
Figure \ref{fig:sky_hist} shows the marginal distributions for the polar
angle from the line-of-sight
between the two LIGO detectors.
This marginal distribution captures the width of the triangulation rings.
All maps are consistent with some differences due to the reconstruction
algorithms. For example, the cWB map has a “Northern Island” near the equator
not seen in other maps. The shape and placement of the island is affected
by the LIGO detector responses at this particular sky location
\cite{GW150914-EMFOLLOW, Essick:2014wwa}.
The Hanford-Livingston network is sensitive to only one polarization through
most of the sky, and cWB uses this to constrain the reconstructed signal,
with the exception of regions like the island where the network sensitivity
is comparable for both polarizations. In this case, cWB relaxes the constraint
and can not break a degeneracy between sky locations near the island. We note
this occurs only when the triangulation ring falls near one of these regions
and may not be present for other events.

To measure the similarity between the skymaps, 
Table II presents the Fidelity $F(p,q) = \sum_i \sqrt{p_i q_i} \in [0, 1]$
for the various algorithms considered, where
$p_i$ and $q_i$ are the probability densities assigned to pixels
at the same coordinates in two different skymaps.
$F$ is closer to one if the maps are more
similar and $F$ closer to zero if the maps are dissimilar.
For comparison, we also include a skymap produced by
LALInference that does not include calibration uncertainties.
This similarity measurement is between 28\% and 87\%  for different pairs of skymaps.
To check the robustness of the parameter estimation results,
we simulated 29 transients with waveforms similar to GW150914, generated
using the SEOBNRv2 approximant  \cite{Taracchini:2013rva, Kumar:2015tha},
by actuating on the mirrors at the end of the 4 km LIGO arms.
We repeat the analysis on each of these hardware injections.
We find similar Fidelity measurements as with the
GW150914 event, suggesting that this level of agreement between the algorithms
is typical for BBH waveforms at the SNR of GW150914.

\begin{table}
    \begin{center}
    \begin{tabular}{c||c|c||c|c|c|c|c|}
               & \multicolumn{2}{c||}{confidence regions} & \multicolumn{4}{c}{Fidelity} \\
               & 50\% & 90\% & LIB & BW  & LALInf & LALNoCE\\ 
        \hline
        \hline
        cWB    &  98 deg$^2$          &  308 deg$^2$          & 0.55  & 0.55 & 0.51 &  0.50 \\
        LIB    & 208 deg$^2$           &  746 deg$^2$          & -      & 0.45 & 0.68 &  0.28 \\
        BW     & 101 deg$^2$          &  634 deg$^2$          & -      &  -    & 0.68  &  0.87 \\
        LALInf & \PESKYFIFTY        & \PESKYNINTY        & -      &  -    & -      &  0.81 \\
        LALNoCE & \PESKYFIFTYNOCALIB & \PESKYNINTYNOCALIB & -      &  -    & -      & -
    \end{tabular}
    \end{center}
        \label{t:fidelity}      
    \caption{Confidence regions and Fidelity values from GW150914. The Fidelity measures the similarity of two skymaps.  The LALInference skymaps are shown both with (LALInf) and without (LALNoCE) calibration uncertainty included.  The shown burst skymaps do not include calibration uncertainties, which would make the uncertainty regions larger.}
\end{table}

\subsection{Waveform reconstruction}

\begin{figure}[h]
\begin{center}
\includegraphics[width=\columnwidth]{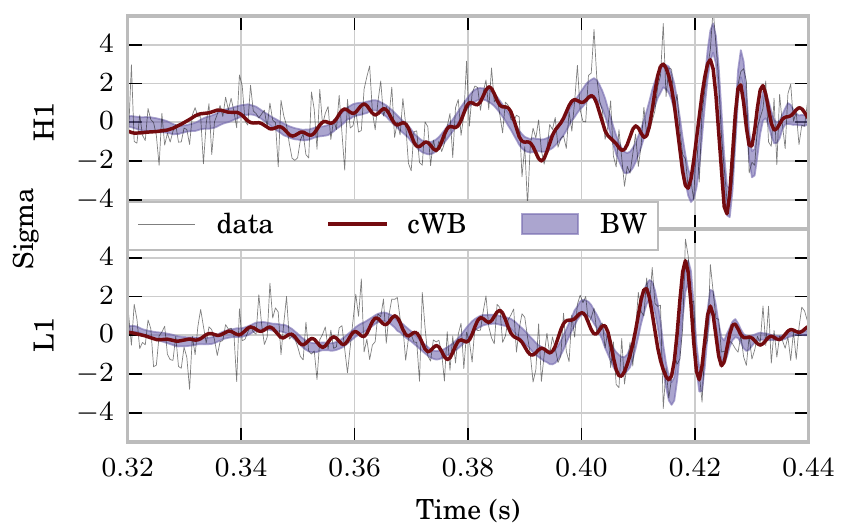}
\end{center}
\vspace*{-0.1in}
\caption{The cWB point estimate for the waveform and the 90\% credible interval from the BayesWave analysis.  The reconstructed waveforms and shown data are
whitened using estimated noise curves for each detector at the
time of the event. On the y-axis, Sigma is a measure of the amplitude in terms of the number of noise standard deviations.\label{fig:BW_CWB}}
\end{figure}

\begin{figure}[h]
\begin{center}
\includegraphics[width=\columnwidth]{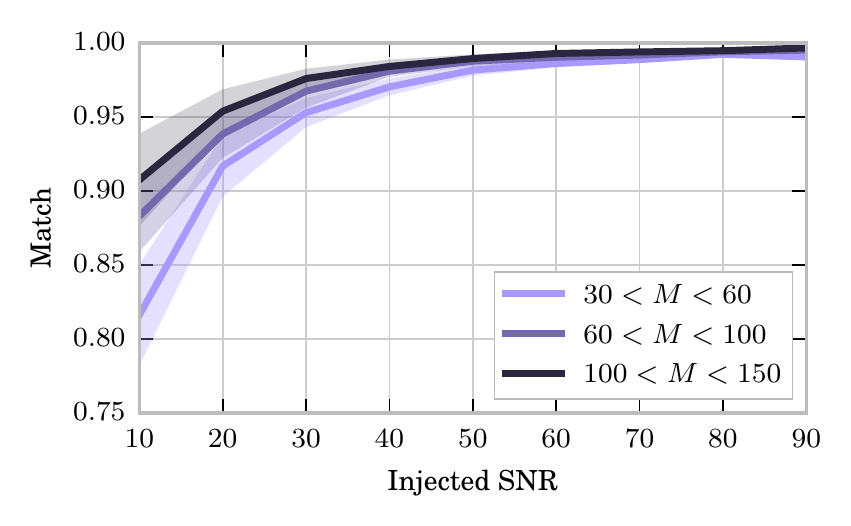}
\end{center}
\vspace*{-0.1in}
\caption{Match between the whitened injected and BayesWave
reconstructed waveforms for the simulation set described in Section
\ref{mdc}.  The line indicates the median match and the shaded region
shows the 1 $\sigma$ uncertainty.  $M$ indicates the total mass of the
black-hole binary, measured in solar masses.
\label{fig:mdc_match}}
\end{figure}

To extract the astrophysical signal from detector noise,
we reconstruct waveforms whose projection
onto both the H1 and L1 detectors is consistent with the data.
The cWB algorithm \cite{klimenko05} performs waveform reconstruction using a
constrained maximum likelihood approach  (See section \ref{cwb_overview}). 
BayesWave \cite{Cornish:2014kda,Littenberg:2014oda} uses a variable dimension continuous wavelet
basis to produce a posterior distribution for the gravitational waveform present in a data set.
In contrast to analyses based on compact object merger templates, 
which attempt to find the best fit parameters
within a well-defined waveform family,
the cWB and BayesWave waveform reconstruction algorithms make very weak assumptions about the form of the signal.
The oLIB pipeline assumes a sine-Gaussian waveform, and so provides a less
detailed reconstruction.
The BayesWave version used in this analysis assumes that the signal is elliptically polarized,
but is otherwise free to reconstruct any astrophysical signal in the searched time-frequency volume.

Figure~\ref{fig:BW_CWB} shows both the cWB point estimate and the BayesWave 90\%
credible interval for the reconstructed, whitened, time-domain signal waveform,
as projected onto each detector.  The waveforms are seen to
largely agree, and include the main expected features from a chirp signal due to a
compact object merger. The BayesWave waveforms have a median match of
\RECONSTRUCTIONOVERLAP 
~with the
posterior samples from a Bayesian analysis that uses waveform templates that account for the inspiral, merger and ringdown phases of the BBH coalescence~\cite{Veitch:2014wba}.

To measure the accuracy of these reconstructions, we use the
set of simulated BBH systems described in Section \ref{mdc}.
For each event recovered by BayesWave, we calculate the match between
the injected and reconstructed waveforms.  The results are shown
in Figure \ref{fig:mdc_match}.
At fixed SNR the match between the simulated and the reconstructed waveform
is systematically higher for higher mass signals because
larger mass BBH signals have a smaller time-frequency volume,
allowing them to be fit with a smaller number of wavelets.
For the simulations similar to
GW150914, in the mass bin from 60 to 100 $\Msun$ and around
network SNR 20, we see most matches are between 90\% and 95\% accurate.

\subsection{Parameter estimation with generic signal features} 
\label{sec:generic_pe}

\begin{figure}[h]
\begin{center}
\includegraphics[width=\columnwidth]{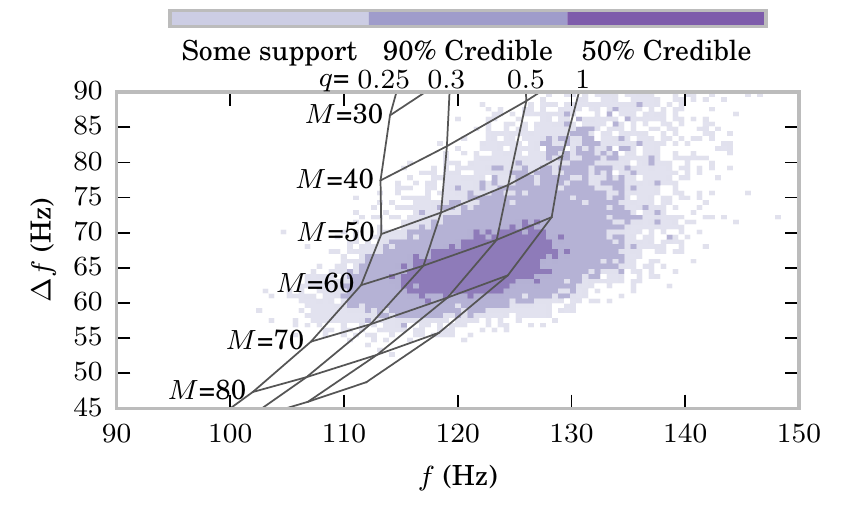}
\end{center}
\vspace*{-0.1in}
\caption{The posterior distributions for the central frequency and bandwidth inferred from the whitened waveform posteriors produced by BayesWave for GW150914 are compared to the values predicted
by the BBH merger templates with zero spin and total mass $M$ (in units of solar mass) and mass ratio $q$, as indicated by the mesh of lines. The regions of high posterior probability are
consistent with the best fit values of total mass and mass ratio ~\cite{GW150914-PARAMESTIM}.}
\label{moments}
\end{figure}

The source parameters of GW150914, such as component masses and spins,
can be well characterized by using an analytical model
of BBH signals to compute their posterior distributions \cite{GW150914-PARAMESTIM}.
Here, we take a different
approach, which uses the outputs of the burst pipelines to provide a coarse estimate of
the model parameters.
The BayesWave and cWB waveform reconstructions can be used to compute a variety of parameters
that summarize the signal, such as the central frequency, duration and
bandwidth. These parameters can then be used to help identify characteristics
of the astrophysical system that generated the signal.
Using waveform templates for a BBH merger, we can derive predictions for
the central frequency and bandwidth of the signal in each
detector as a function of the mass, mass ratio and spins. Figure \ref{moments} shows the
posterior distribution for the central frequency and bandwidth derived from the BayesWave
analysis of GW150914,
with an overlaid grid showing the values predicted from a black-hole merger model with zero
spins and total mass $M$ and mass ratio $q$ as indicated.
From our companion paper,~\cite{GW150914-PARAMESTIM}, the best description of this signal
yields a detector frame total mass of $M =$ \MTOTobsCOMPACT ~$\Msun$
and a mass ratio of $q=$ \MASSRATIOCOMPACT .
Comparing these best fit values to the the regions of high posterior
density shown in Figure \ref{moments}, we find that the
values lie within the 90\% credible interval produced using the
BayesWave outputs.

Applying the same procedure to the 29 GW150914-like hardware injections we found that the central frequency and bandwidth of the
injected signals fell within the 50\% credible interval 50\% of the time, and within the 90\% credible interval 89\% of the time, showing that the
analysis is consistent.

\subsection{Chirp mass from time-frequency signature}
\label{chirpmass}

The cWB pipeline obtains the time-frequency patterns of the events by using
a discrete wavelet transform. Given a pattern with $N$ time-frequency 
components $(t_i,f_i)$, $i=1,...,N$ from a coalescing binary, 
at the leading post-Newtonian order it is described by the time-frequency evolution \cite{PhysRev.136.B1224} 
\begin{equation} 
\frac{96}{5}\pi^{8/3}\left(\frac{\text{G}\mathcal{M}}
{c^3}\right)^{5/3}t+\frac{3}{8}f^{-8/3}+C=0,\label{inteqchirpmassfreq}\end{equation}
where $\mathcal{M}$ is the chirp mass parameter,
G is the gravitational constant, $c$ is the speed of light and $C$ is a 
constant related to the merger time. By fitting this time-frequency evolution 
to the data $(t_i,f_i)$, we can find the mass parameter $\mathcal{M}$~\cite{Tiwari:2015tknm}.
For a signal from a coalescing binary with component masses
$m_1$ and $m_2$ it corresponds to the chirp mass of the system 
$\mathcal{M}=(m_1m_2)^{3/5}/(m_1+m_2)^{1/5}$.
The chirp mass error is estimated using a bootstrapping procedure,
where multiple subsets of data points $(t_i,f_i)$, $i=1,...,N$
are randomly selected to estimate the chirp mass.
  
The real-time search that first detected GW150914 estimated its
detector frame chirp mass to be  $27.6\pm{2.0}\Msun$.
This result is consistent with the LALInference estimate of
\MCobsCOMPACT ~$\Msun$ \cite{GW150914-PARAMESTIM}. 
To check the accuracy of the real-time method, we studied
29 hardware injections with parameters similar to those inferred for GW150914.
We found that this method was able to accurately reconstruct the chirp masses
of these simulated signals, with a precision similar to the quoted uncertainty.

\subsection{Overlap between reconstructed waveform and BBH model}
\label{sec:nrburst}

This section presents the comparison of the reconstructed signal of the event,
from BayesWave and cWB, with predictions from NR.  The goal is to
provide a quantitative check that the recovered signal power is consistent with
a BBH source as predicted by numerical relativity simulations; 
more stringent tests of general relativity are
available in~\cite{GW150914-TESTOFGR}.  By making very weak assumptions about the signal,
the waveform reconstruction provides a largely model-agnostic representation of
the full astrophysical signal content.  In turn, the NR waveform is the direct
solution to the full Einstein equations without any assumptions other than those
necessary to numerically solve the equations, e.g.   finite discretization and
finite extraction radius. 
The NR waveforms used in this study were generated by the code
in~\cite{Pekowsky:2013ska}.  The errors in the phase and amplitude of the
waveform that arise from these approximations are addressed
in~\cite{Hinder:2013oqa}.
Comparing directly to NR waveforms allows us to explore regions of parameter
space where the analytic templates~\cite{GW150914-PARAMESTIM} have not yet been tuned, such as
highly precessing spin configurations and their higher harmonics.  The study is
a simple  way to compare the reconstructed astrophysical signal with the
predictions of general relativity with minimal assumptions.
By comparing the NR waveforms, which cover regions of the parameter space which
are not necessarily well-modelled and include higher harmonics, with the
model-independent reconstructed waveforms which can recover the full astrophysical
signal content, we are sensitive to departures from both the analytic templates used
elsewhere and from the predictions of general relativity.
In fact, we find excellent agreement between
this study and the parameter estimation performed with analytic templates,
as well as with the parameter estimation procedure using only NR waveforms
which is reported in~\cite{NRpe}.  We discuss these findings below.

The natural figure of merit for this comparison is the fitting factor.
We define the network match between the reconstructed waveform 
$s_{\mathrm{rec}}^{(d)}$ in detector $d$ and the NR waveform $h_{\mathrm{NR}}$ by\cite{Allen:2005fk}
\begin{equation}\label{eq:network_match}
     {\cal N} = \frac{\sum_d \max_{t_0, \phi_0} (s^{d}_{\mathrm{rec}}|h_{\mathrm{NR}})_d}
{ \left[\sum_d (h_{\mathrm{NR}}|h_{\mathrm{NR}})_{d} \right]^{1/2} \times \left[\sum_d (s^d_{\mathrm{rec}}|s^d_{\mathrm{rec}})_{d} \right]^{1/2}}.
 \end{equation}
 where the sums run over the H1 and L1 detectors and $(a|b)_d$ defines the
 noise-weighted inner product between waveforms $a$ and $b$ for detector $d$.
The fitting factor is the network match ${\cal N}$ maximized over the
total mass and orbital inclination \cite{Nelder:1965zz}.

The reconstructed waveforms are compared to 102 BBH waveforms that have been
used previously to investigate the feasibility of detecting precession and
higher modes
\cite{2010arXiv1010.5200F,O'Shaughnessy:2010ex,O'Shaughnessy:2012ay,Pekowsky:2013ska,2013PhRvD..87h4008P,2013PhRvD..88b4034H,London:2014cma, jani:2016, galley:2016, schmidt:2016}.
We also include an additional  4 new simulations with intrinsic parameters
motivated by parameter estimation studies of GW150914~\cite{GW150914-PARAMESTIM}.  Note that the
NR simulations are not a continuous representation of the  parameter space, but
rather a discrete set of astrophysically interesting, generic systems.  Each NR
waveform, $h_{\mathrm{NR}}$, is parameterized by the mass ratio
$q=m_2/m_1<1$ and spin configuration of the system.

\begin{figure}[h]
\begin{center}
\includegraphics[width=\columnwidth]{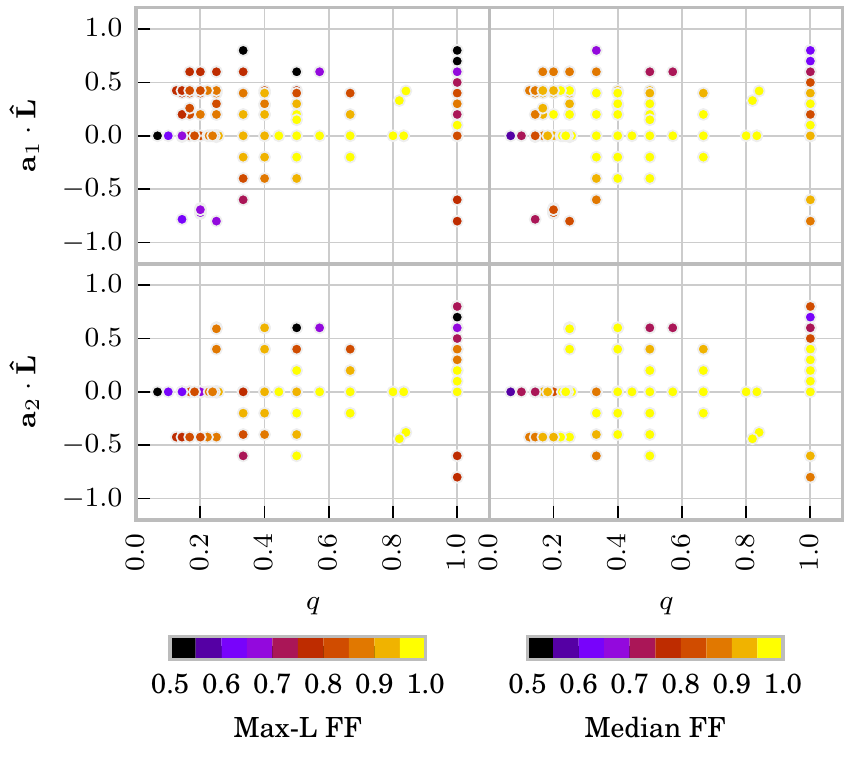}
\end{center}
\vspace*{-0.1in}
\caption{ Fitting factors between cWB (left) and  BayesWave (right) and the NR waveforms,
in terms of the mass ratio $q$ and the dot products between the
component spins and the orbital angular momentum, $\mathbf{a_1}\cdot\mathbf{\hat{L}},\mathbf{a_2}\cdot\mathbf{\hat{L}}$.
The quoted  BayesWave fitting factor values are the median values evaluated
across 1000 posterior waveform samples. \label{fig:BW_FF_s1dotLs2dotL}}
\end{figure}

Figure~\ref{fig:BW_FF_s1dotLs2dotL} shows the fitting factors between 
BayesWave and cWB and the NR waveforms in terms of the mass ratio $q$ and
the dot products between the component spins and the orbital angular momentum,
$\mathbf{a_i}\cdot\mathbf{\hat{L}}$ for $i=1,2$.  The figure also serves to demonstrate the
coverage of the parameter space by the NR simulations.  We find that the
parameter space of NR waveforms favored by both algorithms is similar.
Specifically, nearly symmetric mass configurations and small values for
$\mathbf{a_i}\cdot\mathbf{\hat{L}}$ for both components are preferred, although
the lack of variation in the fitting factor across the spin-space suggests this
is not strongly constrained.  

The BayesWave and cWB reconstructed waveforms have a fitting factor with the
best fit NR waveform of 0.95 and 0.87, respectively. Fits within 1\% of the
best fit value are achieved with detector frame total mass in the range
66.4 -- 74.8 $\Msun$ for BayesWave and 67.9 -- 75.7 $\Msun$ for cWB. This is in
excellent agreement with the range \MTOTobsRANGE~$\Msun$ estimated using
LALInference \cite{GW150914-PARAMESTIM}. The chirp mass of NR waveforms within 1\% of the best
fit to the BayesWave and cWB reconstructions is in the range 27.4 -- 32.6
$\Msun$ and 27.8 -- 33.0 $\Msun$, again with close overlap to the LALInference
result of \MCobsRANGE ~$\Msun$.

In addition to matching parameter estimation performed using analytic waveform models
in~\cite{GW150914-PARAMESTIM},
the parameter bounds shown here
are consistent with those obtained via the time-frequency analyses in
sections~\ref{sec:generic_pe} and \ref{chirpmass}.
Findings similar to those here are reported
in~\cite{NRpe} where a suite of NR waveforms, including those used in this
study, are compared directly with the data in a novel Bayesian analysis.  Again,
the parameter space preferred by that study clearly overlaps with that here.
The agreement between the
analytic waveform results and the Bayesian NR analysis helps to validate the use
of those waveform templates.  Meanwhile, the overlap with the model-independent
reconstructions here demonstrates that there is no significant additional signal
content which the NR waveforms fail to represent, as would be the case for
sources other than BBH.  The concordance between the findings from these
three studies further serves to highlight the BBH origin of GW150914.

\section{Discussion}


All-sky searches for short-duration gravitational wave
bursts scan a broad parameter space to identify the
presence of gravitational wave signals in the data.
They discovered GW150914 in a low-latency online analysis, 
and identified it as clearly distinct from detector
noise events.
Further analysis of GW150914
showed that the reconstructed waveform of the signal is consistent with expectations for
a binary black hole merger. 
Outputs of the burst pipelines
were also used to estimate the mass parameters of the source,
in agreement with more specialized techniques.

The discovery of GW150914 is a turning-point in
gravitational wave astronomy. 
At the time of the discovery, low-latency burst searches
were configured to search a broad parameter space, similar
to gravitational-wave burst searches performed during the
initial detector era.
The large search parameter space was seen to overlap with
high-mass binary black hole signals in studies with simulated data,
an observation confirmed by the detection of GW150914.
Looking towards the future, the emphasis on searches with minimal assumptions of the
waveform morphology allows for gravitational wave burst
searches to explore the vast discovery space of gravitational
wave transients from a variety of potential sources.

Beyond the challenge of detecting gravitational waves,
burst parameter estimation tools, which make weak signal assumptions,
have demonstrated their ability to extract astrophysical information about the
progenitor of GW150914.
Rapid sky localization of transient sources will facilitate multi-messenger astronomy
and allow for improved characterization of gravitational wave signal progenitors.
Many of the tools used for GW150914, such as waveform reconstruction, 
have applications beyond gravitational waves from binary coalescences. 

The methods described in this work will also be used to search the full
data set from the first observing run in the advanced detector era and beyond.
Gravitational wave burst searches, through their detection and analysis of GW150914
have shown, that they are ready to contribute to an era of gravitational wave astronomy.


\bigskip\noindent\textit{Acknowledgments} ---
 The authors gratefully acknowledge the support of the United States
National Science Foundation (NSF) for the construction and operation of the
LIGO Laboratory and Advanced LIGO as well as the Science and Technology Facilities Council (STFC) of the
United Kingdom, the Max-Planck-Society (MPS), and the State of
Niedersachsen/Germany for support of the construction of Advanced LIGO 
and construction and operation of the GEO\,600 detector. 
Additional support for Advanced LIGO was provided by the Australian Research Council.
The authors gratefully acknowledge the Italian Istituto Nazionale di Fisica Nucleare (INFN),  
the French Centre National de la Recherche Scientifique (CNRS) and
the Foundation for Fundamental Research on Matter supported by the Netherlands Organisation for Scientific Research, 
for the construction and operation of the Virgo detector
and the creation and support  of the EGO consortium. 
The authors also gratefully acknowledge research support from these agencies as well as by 
the Council of Scientific and Industrial Research of India, 
Department of Science and Technology, India,
Science \& Engineering Research Board (SERB), India,
Ministry of Human Resource Development, India,
the Spanish Ministerio de Econom\'ia y Competitividad,
the Conselleria d'Economia i Competitivitat and Conselleria d'Educaci\'o, Cultura i Universitats of the Govern de les Illes Balears,
the National Science Centre of Poland,
the European Commission,
the Royal Society, 
the Scottish Funding Council, 
the Scottish Universities Physics Alliance, 
the Hungarian Scientific Research Fund (OTKA),
the Lyon Institute of Origins (LIO),
the National Research Foundation of Korea,
Industry Canada and the Province of Ontario through the Ministry of Economic Development and Innovation, 
the National Science and Engineering Research Council Canada,
Canadian Institute for Advanced Research,
the Brazilian Ministry of Science, Technology, and Innovation,
Russian Foundation for Basic Research,
the Leverhulme Trust, 
the Research Corporation, 
Ministry of Science and Technology (MOST), Taiwan
and
the Kavli Foundation.
The authors gratefully acknowledge the support of the NSF, STFC, MPS, INFN, CNRS and the
State of Niedersachsen/Germany for provision of computational resources.

This  article  has  been  assigned the document numbers LIGO-P1500229.

\bibliographystyle{unsrt}
\bibliography{GW150914_burst,GW150914_refs_arxiv.bib}

\end{document}